\documentclass[%
 preprint,
 amsmath,amssymb,
]{revtex4-1}

\usepackage{graphicx}
\usepackage{dcolumn}
\usepackage{bm}
\usepackage{hyperref}
\usepackage{amsmath}

\usepackage[utf8]{inputenc}
\usepackage{booktabs}

\usepackage{color}
\definecolor{redcolor}{rgb}{1.0,0.,0.}

\begin{document}

\preprint{APS/123-QED}

\title{A complex network approach to political analysis: application to the Brazilian Chamber of Deputies}

\author{Ana C. M. Brito}
\affiliation{%
Institute of Mathematics and Computer Science, University of S\~{a}o Paulo,
S\~{a}o Carlos, SP,  Brazil
}%


\author{Filipi N. Silva}
\affiliation{
Indiana University Network Science Institute, Bloomington, Indiana 47408, USA
}%

\author{Diego R. Amancio}
\email{diego@icmc.usp.br}
\affiliation{%
Institute of Mathematics and Computer Science, Department of Computer Science, University of S\~{a}o Paulo,
S\~{a}o Carlos, SP,  Brazil
}%

\date{\today}

\begin{abstract}
In this paper, we introduce a network-based methodology to study how clusters represented by political entities evolve over time. We constructed networks of voting data from the Brazilian Chamber of Deputies, where deputies are nodes and edges are represented by voting similarity among deputies. The Brazilian Chamber of deputies is characterized by a multi-party political system. Thus, we would expect a broad spectrum of ideas to be represented. Our results, however, revealed that plurality of ideas is not present at all: the effective number of communities representing ideas based on agreement/disagreement in propositions is about 3 over the entire studied time span. The obtained results also revealed different patterns of coalitions between distinct parties. Finally, we also found signs of early party isolation before presidential impeachment proceedings effectively started. We believe that the proposed framework could be used to complement the study of political dynamics and even applied in similar social networks where individuals are organized in a complex manner.
%



\end{abstract}

\maketitle


\section{Introduction}
In recent years, the availability of information in the form of open datasets together with the capabilities to store and process data have been promoting the rise of many new studies in a variety of disciplines, including Biology~\cite{huang2008bioinformatics}, Social Sciences~\cite{java2007we}, Linguistics~\cite{acerbi2013expression,he2011self,correa2018word,michel2011quantitative} and  Physics~\cite{baldi2014searching,broido2019scale}. Many of these systems can be regarded as being too complex for traditional methodologies in which each of its components is isolated and studied individually. This includes the political dynamics of a country, which may depends on many aspects such as economic factors, culture, mass media, social media, etc. In a democratic system, these aspects are reflected (or should be) on the decisions made by people's representatives and on how they are organized (such as partisanship or alliances between parties). In this context, the relationships between politicians could be drawn indirectly from how they vote in propositions~\cite{moody_mucha_2013,dal2014voting,cherepnalkoski2016cohesion}, i.e., their agreement or disagreement among a set of voted proposals.

Politicians with similar voting patterns can be understood as having similar views and interests, thus can be connected in a political network. Being a complex system, it makes sense to explore such a system by using methods borrowed from network science~\cite{newman2018networks}. The Brazilian political environment becomes an interesting system to be studied under this approach since it is a multi-party system, thus, supposedly presenting a diverse spectrum of political ideas. Also, it is a system that has undergone many changes over the past few years. Moreover, data for decisions and political organization are openly available.

In this paper, we propose a framework based on complex networks to study the evolution of the Brazilian political system in terms of how politicians vote in proposals along time. For that, we collected and processed voting data from 1991 to 2018, spanning 6 terms in the lower chamber and build vote correlation networks. Because the obtained networks were weighted and dense, we applied a filtering method to preserve the community structure. Using concepts borrowed from network science, we defined relevant political concepts, including \emph{coalition}, \emph{fragmentation} and \emph{isolation}. We also devised a measure to quantify the \emph{effective} number of groups, regardless of the number of political parties.

Several interesting results were obtained in the Brazilian political scenario. We found in most of  years a low correspondence between communities and topological groups, meaning that the majority of deputies tend to share votes with different parties. Most importantly, despite the large number of political parties in the Brazilian political scenario (29 parties in 2019), effectively only roughly $3$ groups of deputies were identified. The quantification and visualization of shared coalitions and isolation were consistent with the political scenario unfolded along the considered period. We also found different behavior of coalitions: while some parties tend to be aligned to the president party, other parties are always in distinct network communities. Finally, we identified an isolation pattern that could be an early indication of the start of an impeachment process. While we focused on the analysis of the Brazilian case, we advocate that the methods are robust enough to be applied in other democratic systems and even in other social networks.

The paper is structured as follows. In Section \ref{sec:related}, we review some recent works using a graph representation to study political systems. In Section \ref{submet1}, we detail the creation of networks from voting data. The procedure employed to prune the obtained weighted, dense networks is presented in Section \ref{submet2}. Section \ref{submet3} presents a discussion on how to effectively measure the quality and quantity of clusters identified in networks. In Section \ref{submet:end}, we introduce some measurements to quantify the concepts of political coalition, isolation and fragmentation. Finally, results and future works are discussed in Sections \ref{sec:results} and \ref{sec:conc}, respectively.






\section{Related works} \label{sec:related}




\label{sub-sec:redesvotacao}

Some studies used the concept of complex networks to model some political concepts, including the concept of political \emph{coalition} and \emph{cohesion}~\cite{dal2014voting,cherepnalkoski2016cohesion,levorato2016brazilian, ferreira2018analyzing}. Both concepts are quantified via networks extracted from  parliamentary votes.

In~\cite{dal2014voting}, network edges are established according to the similarity of votes between parliamentarians. Votes were considered to be chosen among three options. Deputies can vote for and against a specific proposition. In addition, they may abstain from voting. All data were extracted from the Italian parliament in 2013.
A topological analysis to quantify the concepts of \emph{cohesion}, \emph{polarization} and \emph{government influence} was created in terms of network parameters. The cohesion of political parties was defined using the following two density indexes:
\begin{equation}
    \rho_{int}(C) = \frac{1}{{n_c(n_c-1)}} \sum_{i \in C} \sum_{j \in C} w_{ij},
    \label{eq:intra}
\end{equation}
\begin{equation}
    \rho_{ext}(C) = \frac{1}{n_c(n_c-1)} \sum_{i\in C} \sum_{j \not\in C} w_{ij},
    \label{eq:inter}
\end{equation}
where $n_c$ is the number of node in cluster (community) $C$, and $w_{ij}$ is the weight of the edge linking nodes $i$ and $j$. A given community $C$ is considered cohesive if the internal density $\rho_{int}(C)$ is significantly higher than the average internal density of the whole network ($\langle \rho_{int} \rangle$). In a similar fashion, $C$ may also be considered cohesive whenever $\rho_{ext}(C) > \langle \rho_{ext} \rangle$.

In~\cite{dal2014voting}, network communities and relevant parliamentarians are identified by measuring the effect of removing and inserting nodes in distinct network communities. More specifically, the authors measured the variation in the modularity $\textrm{d}Q$~\cite{de2013community} when parliamentarians with distinct political positions are placed in their respective opposite political communities. The degree of polarization was then defined as proportional to the absolute variation $|\textrm{d}Q|$ resulting from this procedure.

The cohesion defined in equations
\ref{eq:intra} and \ref{eq:inter} revealed that some parties are well defined, while some parliamentarians display a voting pattern that is most similar to the patterns observed in other parties. Upon analyzing the modularity of the networks, the authors advocate that it may not be the best alternative to characterize the obtained clustered networks.  Their analysis also showed that the emergence of three main clusters, where two groups represent parliamentarians voting according to the government preferences, while the remaining cluster was found to represent opposition parties. Finally, a temporal analysis revealed patterns of voting behavior along time. Such a temporal analysis was useful to identify e.g. parties do not display a consistent voting behavior.

In~\cite{levorato2016brazilian}, a dataset of propositions analyzed between 2011 and 2016 was used to analyze votes in the Brazilian political scenario. Nodes were defined as deputies, which are linked by the number of identical votes in the considered period. Different variations in the network construction were considered because abstentions may lead to different interpretations on whether two deputies agree when at least one of them abstain from voting. This study also defined political concepts such as polarization in social networks using correlation clustering~\cite{becker2005survey} and its symmetric relaxed version to quantify polarization.

The analysis carried out in~\cite{levorato2016brazilian} revealed that several parties do not comply with their respective original ideas as expressed during the campaign period. In addition, this study showed that, in recent years, the group of deputies allied to the Brazilian government diminished. Even after presidential reelection, allied deputies turned out to display a voting pattern in disagreement with government propositions. Finally, this study reports that the distance between centre- and right-wing deputies decreased in recent years.

The research conducted in~\cite{ferreira2018analyzing} studied the network community organization of parliamentarians, focusing on their temporal evolution. The data were obtained from both Brazilian Chamber of Deputies and US Parliament from 2003 to 2017.
Temporal networks were created without overlapping. In the network representation, parliamentarians are nodes and edge weights are established according to the vote similarity. In the Brazilian case,   considered two additional possibilities for each vote: (i) abstention; and (ii) obstruction. In the latter, it may also be considered as a vote against the proposition under discussion. The authors proposed a measurement that depends on the community structure of the network. The \emph{Partisan Discipline}, computed for a given parliamentarian $m$, was defined as
\begin{equation}
    \textrm{pd}(m) = \frac{1}{N}\sum_{i=1}^N I(m,p_m,i),
    \label{eq:pdm}
\end{equation}
where $N$ is the total number of propositions analyzed and $I(m,p_m,i) = 1$ iff $m$ and the respective group (i.e. network community or party) to which he/she belongs voted equally in the $i$-th proposition.

The application of equation \ref{eq:pdm} for the data obtained from the Brazilian Chamber of Deputies revealed that the values of \emph{Partisan Discipline} are higher when groups are the identified communities, in comparison to groups being traditional parties. For the US Parliament, the obtained values of \emph{Partisan Discipline} are similar when groups and communities are considered. The concept of polarization was analyzed by observing that cohesive communities emerges in a political scenario characterized by a high degree of polarization. Because  members of the same community should have similar sets of neighbors, the number of shared neighbors was used as an indicative of political polarization. The polarization analysis revealed that the Brazilian Chamber of Deputies  has a low degree of polarization. In addition, this study showed that deputies oftentimes do not stay in the same dense, well-defined community for long periods.

Some studies used political networks to identify the ideology of political parties.
While many studies focused on voting data, the study reported in~\cite{faustino2019} focused on party-switching affiliations from one election to another to generate networks. In such a network, nodes are parties and two parties are linked if there was a migration of deputies between these parties. The community structure of the obtained networks revealed a tendency of deputies to switch between parties with similar ideology. A model was also created to quantify the ideology of political parties on the left-right scale. The methodology was found to be robust as it does not depend on the level of representation of parties, since only affiliation data is required to create political networks.

The European Parliament was studied using networks in the work conducted in~\cite{cherepnalkoski2016cohesion}. A dataset with roll-call voting data from the European Parliament was used to create a network where parliamentarians are nodes and edges are weights according to the similarity of voting patterns. Twitter~\footnote{\url{https://twitter.com}}  was also used to create a network where parliamentarians are connected if there is a ``retweet'' relationship between them. The main results revealed the emergence of political coalitions, characterized by cooperation of opposite political wing groups. In addition, they found a tendency of cooperation among small and large political groups. Finally, this study also reports that political alignments in Twitter are consistent with voting patterns.

\textcolor{black}{Unlike the above mentioned studies, we propose metrics based on shortest path lengths which can be understood as the connection strength between parliament members. We also use temporal series to show our results, enabling thus the visualization of cluster dynamics. Differently from other works, we compare the traditional structure of political parties with the actual structure observed from voting behavior. }
Our focus is on the characterization of groups, which encompasses the definition of political fragmentation, isolation and diversity of parties.

\section{Methodology}

The proposed \textit{framework} to analyze the relationship among deputies according to their voting patterns can be summarized in the following steps:

\begin{figure}
    \centering
    \includegraphics[width=0.96\textwidth]{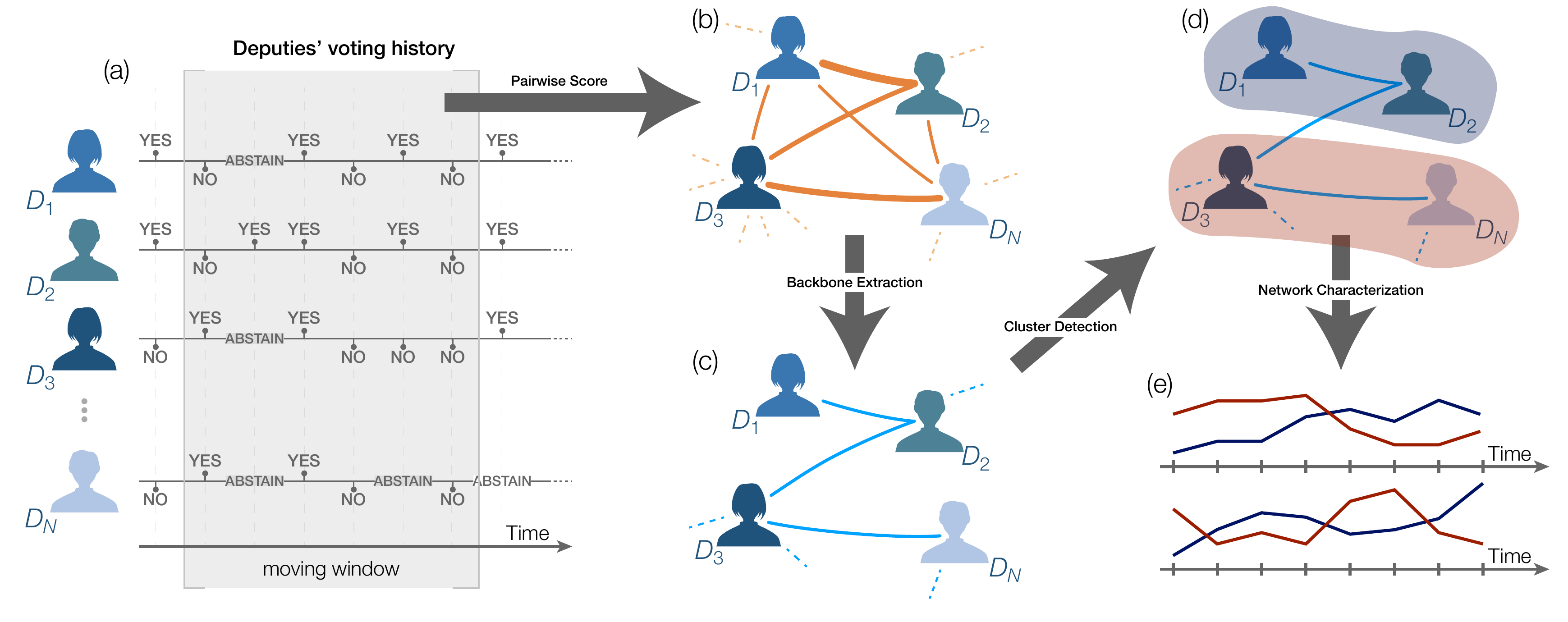}
    \caption{Schematic representing the methodology in this paper. In (a), the voting data is analyzed. In (b), the pairwise similarity of votes in distinct propositions generates a weighted network of deputies. Edges weight represent the distance between two deputies, according to their voting behavior. In (c), edges are filtered so as to preserve the community structure. In (d), we identify the community structure of the network.
    Finally, in (e), the resulting network is analyzed by measuring e.g. distance between parties or clusters of deputies.}
    \label{fig:schematics}
\end{figure}

\begin{enumerate}
    \item \emph{Network construction}: a network is created from voting patterns in a given period. Two deputies are linked if they vote in a similar fashion in several propositions.

    \item \emph{Network backbone extraction}: this step involves the removal of the weakest edges. This is essential to create a network that can be treated by most of the traditional community detection methods. Mostly important, the removal of the weakest edges does not affect the community structure of the obtained networks.

    \item \emph{Cluster detection and analysis}: community detection methods are employed to identify groups of deputies displaying similar voting patterns.

    \item \emph{Network characterization}: metrics are extracted from the network in order to quantify political concepts such as fragmentation, isolation and effective number of groups.

\end{enumerate}
These four steps are detailed, respectively, in subsections \ref{submet1}, \ref{submet2}, \ref{submet3} and \ref{submet:end}. All involved steps are illustrated in Figure \ref{fig:schematics}.

\subsection{Network construction} \label{submet1}

Networks are constructed by representing deputies as nodes. Edges are established between two deputies if the percentage of agreement considering all propositions in the dataset is higher than the percentage of disagreement. Mathematically, all votes established by the $i$-th deputy can be stored in a $N$-dimensional vector $\mathbf{v}^{(i)}$, where $N$ is the total number of considered propositions. The $j$-th element of $\mathbf{v}$ may store three different values. $\mathbf{v}_j$ may take the values $-1$ or $+1$, representing respectively a vote for or against a given proposition. 
Thus, the pairwise level of agreement between two deputies is defined as:
\begin{equation} \label{eq:sim}
    w_{ij} = \frac{1}{N} \sum_{k=1}^{N} \mathbf{v}^{(i)}_k \mathbf{v}^{(j)}_k = \frac{\mathbf{v}^{(i)} \cdot \mathbf{v}^{(j)}}{N} ,
\end{equation}
where $\mathbf{v}^{(i)} \cdot \mathbf{v}^{(j)}$ is the dot product between $\mathbf{v}^{(i)}$ and $\mathbf{v}^{(j)}$. Note that, when deputies have the same opinion about the proposition being voted, the level of agreement is increased by 1. Conversely, when views are opposite, $w_{ij}$ is decreased by 1. Whenever at least one of the deputies absence from voting in a proposition, that proposition is not considered when computing $w_{ij}$.

Because edges weights are given by $w_{ij}$, all obtained networks are undirected and weighted. As we shall show, the networks may be constructed using a particular time interval, and also considering sliding (overlapping) windows of fixed length.

\subsection{Network backbone extraction} \label{submet2}


Even considering only agreement between deputies as connections, the proposed approach can still lead to densely connected networks. While such networks can still be regarded as a weighted, it is usually desirable to remove noise from data, such as weak connections that may correspond to spurious links. For that reason, we included a preprocessing step to our analysis pipeline in which we deal with the problem of keeping only important connections in a network. Such a task is known as edge pruning~\cite{zhou2012simplification}.

A simple way to accomplish edge pruning is keeping only edges with weights larger than a fixed threshold. Such a threshold can be determined, for instance, in terms of percentiles to reach a certain network density. The main problem with that approach is that it may introduce some bias toward highly connected nodes which can lead to substantial undesired effects on the network topology.

A more sophisticated approach to the edge pruning task is the backbone extraction method \cite{serrano2009extracting}. This technique is based on the idea that when determining the importance of an edge, we must take into account how its ending points are connected to the network. More specifically, the authors proposed that the importance of a node can be determined in terms of a disparity filter, so that a value $\alpha_{ij}$ for an edge can be drawn from a $p$-value determined in terms of null model. This null model considers the probability of a node having an edge with a certain weight based on its other connections. In practice, one can compute $\alpha_{ij}$ for a certain edge $e_{ij}$~
existing in the network as:
\begin{equation} \label{eq:alpha}
\alpha_{ij} = 1 - (k_i-1)\int_{0}^{p_{ij}}(1-x)^{k_i-2} dx ,
\end{equation}
where $k_i$ is the number of connections for node $i$ and $p_{ij}$ is a probability defined as
\begin{equation} \label{eq:prob}
p_{ij} = {w_{ij} \over \sum_{{ik}\,\in\,E}w_{ik}}.
\end{equation}
The resulting network is obtained by removing all edges with $\alpha_{ij}$ higher than a certain threshold $\alpha$. Because the networks under analysis are undirected, we consider, for each edge $e_{ij}$, the lowest between $\alpha_{ij}$ and $\alpha_{ji}$.
Here, we opted no to use a fixed threshold for $\alpha$. We instead used the smallest $\alpha$ such that each network have at least $80\%$ of the nodes belonging to the giant component. This results in more consistent networks along time and allow us to analyze the data based on the major component. This is important in network analysis because some measurements, such as shortest path distances, are only measured in a connected component.
%

\begin{figure}
    \centering
    \includegraphics[width=1\linewidth]{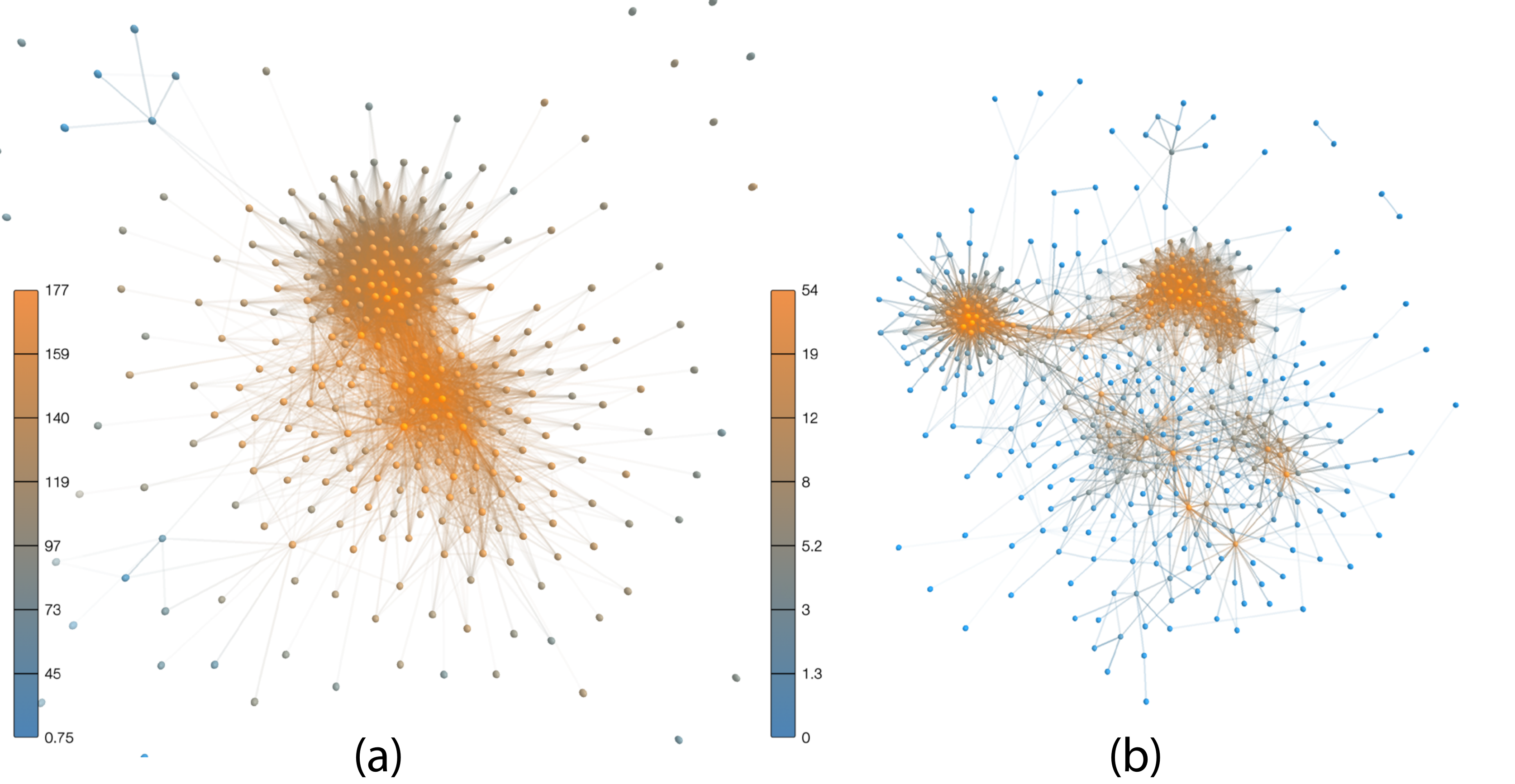}
    \caption{Network obtained considered all proposition voted in 2012. This network was constructed using the procedure described in (a)  Section \ref{submet1} (before edge pruning);  and (b)  Section \ref{submet2} (after edge pruning). Each node represents a member from the Brazilian lower chamber. Two members are linked if they voted in a similar way. Colors represent nodes degree.  This visualization was created using the \emph{Networks3D} visualizer software~\cite{silva2016using}.\label{fig:visualization}}
    \label{fig:v1998f}
\end{figure}

%
In order to illustrate the edge pruning process, we show in Figure \ref{fig:v1998f} a visualization of a network before and after the pruning process is applied. Interestingly, there is a clear separation in  communities of deputies with very distinct voting patterns. The emergence of several communities is not as clear when the filtering procedure is not applied.
%
%

\subsection{Cluster detection and analysis} \label{submet3}

Once the network is obtained, the clustering of nodes is necessary to understand how deputies are clustered by taking into account voting patterns. The identification of communities is also important to analyze if the organization of deputies in parties is consistent with the natural organization emerging from their voting behavior.

Several community detection methods rely on the \emph{modularity} to measure the quality of the obtained clusters in complex networks~\cite{blondel2008fast,ribeiro2018dynamical,de2016using}. This measurement quantifies the number of links that occur inside a community that are above the same number expected in a null model. Mathematically, the network modularity $Q$ is given by
\begin{equation}
    Q = \frac{1}{2m} \sum_{ij} \Bigg{[} a_{ij} - \frac{k_i k_j}{2m} \Bigg{]} \delta(c_i,c_j),
\end{equation}
where $a_{ij} = 1$ if nodes $i$ and $j$ are connected and  $a_{ij} = 0$, otherwise. $m = 1/2 \sum_{ij} a_{ij}$ is the total number of edges in the network and $k_i = \sum_j a_{ij}$.
$\delta(c_i,c_j)=1$ only if nodes $i$ and $j$ are in the same community. Otherwise, $\delta(c_i,c_j)=0$.

In this particular study, we used the Leiden method~\cite{traag2019louvain}, which is an improved version of the Louvain algorithm~\cite{blondel2008fast}. While we report our results based on this Leiden method, a preliminary study revealed that the choice of community detection method has no influence on the results reported in this paper.

The Louvain algorithm is based on the iterative repetition of two steps. First, each single node is considered as a single community in the network. Then, for each node, the method evaluates the gain in modularity when it is removed from its current community and placed to a neighbor community. The change of community membership is effectively performed if such a change corresponds to the highest gain in modularity among all possible changes in the current iteration. If the highest gain is negative (i.e. the change of community membership leads to a decrease in modularity), then the node stays in its original community. This process is repeated until no gain in modularity is achieved. The efficiency of the method stems from the fact that the gain of modularity can be computed in a efficient way. The algorithm proposed in~\cite{traag2019louvain} guarantees that the obtained communities are connected.

One particular feature that we are interested in the study of political networks is the correspondence between political parties and network communities. Such an analysis can be performed by comparing the total number of clusters obtained from these two partitioning strategies. A more refined notion group diversity (i.e. total number of groups) in a partition  -- obtained via political parties or community detection -- may be derived from the concept of \emph{true diversity}~\cite{hill1973diversity}. This quantity measures the effective number of groups in a partition by considering the size of groups. Let $\mathcal{S} = \{s_1,s_2,\ldots s_R\}$ be the size of each group in a partition comprising $R$ groups. The computation of the true diversity requires that the sizes in $\mathcal{S}$ are normalized. For this reason, they are normalized to yield the following normalized set of normalized sizes $\Pi = \{\pi_1,\pi_2,\ldots \pi_R\}$, where $\pi_i = s_i / \sum_j s_j$. The true diversity $D^{(q)}$ of the distribution $\Pi$ is defined as
\begin{equation}
    D^{(q)} = \Bigg{(} \sum_{i=1}^R \pi_i^q \Bigg{)}^{1/(1-q)},
\end{equation}
where $q$ is the exponent given to the weighted generalized mean of $\pi_i$'s. If $q=1$, the true diversity can be computed as
\begin{equation} \label{eq:dive}
    D^{(q=1)} = \lim_{q\to 1} \Bigg{(} \sum_{i=1}^R \pi_i^q \Bigg{)}^{1/(1-q)} = \frac{1}{\prod_{i=1}^R \pi_i^{\pi_i}} = \exp \Bigg{(} - \sum_{i=1}^R \pi_i \ln \pi_i \Bigg{)},
\end{equation}
which corresponds to the exponential of the entropy of $\Pi$. The true diversity has been employed in other contexts to measure the \emph{effective} number of species~\cite{chao2016phylogenetic}. In network theory, the accessibility of nodes in complex networks is measured in terms of the true diversity and quantify the \emph{effective} number of neighbors~\cite{correa2017patterns,de2017knowledge,2015comparing,travenccolo2008accessibility,TOHALINO2018526}. Because we are measuring the diversity of clusters size, hereafter we refer to the true diversity as the \emph{effective} number of political parties and communities.

\subsection{Network characterization} \label{submet:end}

In this section, we introduce the measurements that are used to analyze network obtained from voting patterns. We are specially interested in the analysis of distances among parties along time. The first important measurement that can be extracted from political networks is the topological distance $d(A,B)$ between two groups $A$ and $B$, defined as the average distance among their elements:
\begin{equation} \label{eq:distab}
    d(A,B) = \frac{1}{|A \times B|} \sum_{(a,b) \in A \times B} l(a,b).
\end{equation}
where $l(a,b)$ is the shortest path length between $a \in A$ and $b \in B$. When groups are political parties, the distance $d(A,B)$ can be used to quantify the level of \emph{coalition} between $A$ and $B$. In this case, low values of distances correspond to strong coalitions.

Another important quantity is the average distance $d_G$ between all nodes in the network. It is defined as:
\begin{equation} \label{eq:distglob}
    d_G = \frac{2}{n(n-1)} \sum_{a \neq b} l(a,b).
\end{equation}
Note that the distance $l$ can not be computed directly from the network obtained as described in Section \ref{submet2}, because edge weights represent a similarity relationship. In order to map the similarity $w_{ij}$ defined in equation \ref{eq:sim} into a dissimilarity index, we used the following equation:
\begin{equation}
    \Delta(w_{ij}) = \big{[}2\cdot(1-w_{ij})\big{]}^{1/2}.
    \label{eq:dist}
\end{equation}
A more detailed description on network transformations, such as the one in equation \ref{eq:dist}, can be found in~\cite{comin2016complex}.
%
%


The political \emph{isolation} of group $A$, $I(A)$, is defined as the distance between group $A$ and all other groups in a given division of the network.
It is defined as
\begin{equation}
    I(A) =  \sum_{X \neq A} |X| \cdot d(A,X) ~ \ \big{/} \sum_{X \neq A} |X|,
\label{eq:iso}
\end{equation}
which represents the average distance (as defined in equation \ref{eq:distab}) between $A$ and all other groups. Note that the distances are weighted by the size of each group $X \neq A$. According to equation \ref{eq:iso}, the isolation reflects how distant nodes belonging to the group of reference are from the other nodes in the network. Interestingly, there is an analogy between the isolation of groups and the ability (complexity) of unsupervised classification~\cite{rodriguez2019clustering}. In other words, isolated groups are analogous to outlier clusters in unsupervised data analysis.

The concept of political \emph{fragmentation} for group $A$, $F(A)$, can be defined in terms of intra-group dispersion, i.e.
\begin{equation} \label{eq:frag}
    F(A) = d(A,A).
\end{equation}
Therefore, a group is highly fragmented whenever the nodes in that group are distant from each other, according to the topology of the network. Note that the quantities defined in equations \ref{eq:iso} and \ref{eq:frag} are independent. High-values of isolation do not imply low values of fragmentation and vice-versa.
%




\section{Dataset}









The voting data were obtained from the Brazilian Chamber of Deputies~\footnote{\url{https://dadosabertos.camara.leg.br.} Last accessed: September, 2019.}. The dataset comprises a total of $3,530$ propositions voted between April, 1991 and Juny, 2019.
The most frequent political parties appearing in this dataset are: Brazilian Democratic Movement  (\emph{Movimento Democrático Brasileiro} - MDB), Brazilian Social Democracy Party (\emph{Partido da Social Democracia Brasileira} - PSDB), Workers' Party (\emph{Partido dos Trabalhadores} - PT), Progressive Party (\emph{Partido Progressista} - PP), Democrats (\emph{Democratas} - DEM), Communist Party of Brazil (\emph{Partido Comunista do Brasil} - PCdoB),  Brazilian Socialist Party (\emph{Partido Socialista Brasileiro} - PSB), Brazilian Labour Party (\emph{Partido Trabalhista Brasileiro} - PTB), Liberal Party (\emph{Partido Liberal} - PL), Democratic Labour Party (\emph{Partido Democrático Trabalhista} - PDT) and Social Liberal Party (\emph{Partido Social Liberal} - PSL). It is important to note that a few of these parties have undergone splitting or merging processes leading to changes in their names over the past few years. However, for the sake of simplicity, we only consider their most recent names, as of 2019.

\begin{figure}[hbtp!]
    \centering
    \includegraphics[width=\linewidth]{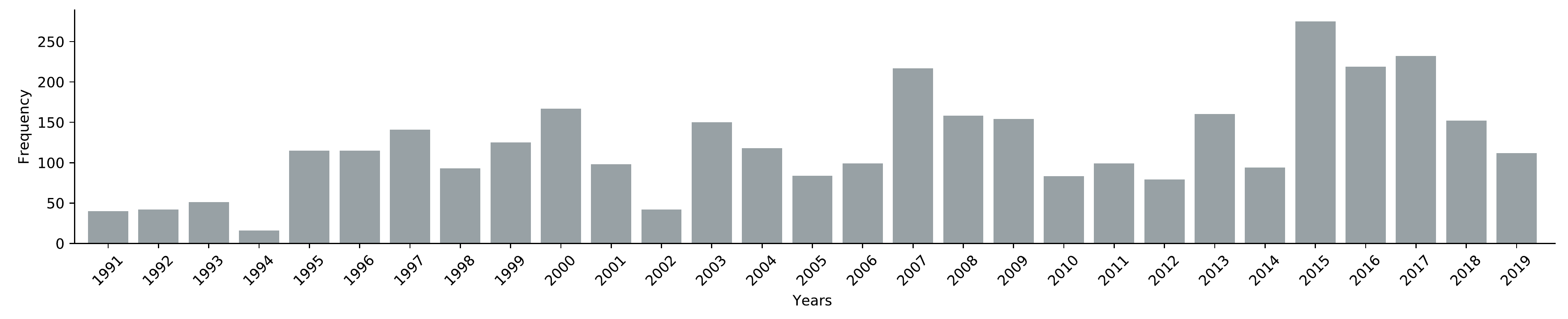}
    \caption{
    Yearly number of propositions analyzed by the Brazilian Chamber of Deputies between 1991 and 2019. While the number of propositions analyzed along time is not regular, more recently at least $100$ propositions were analyzed in each year. }
    \label{fig:props}
\end{figure}

\begin{figure}[hbtp!]
    \centering
    \includegraphics[width=\linewidth]{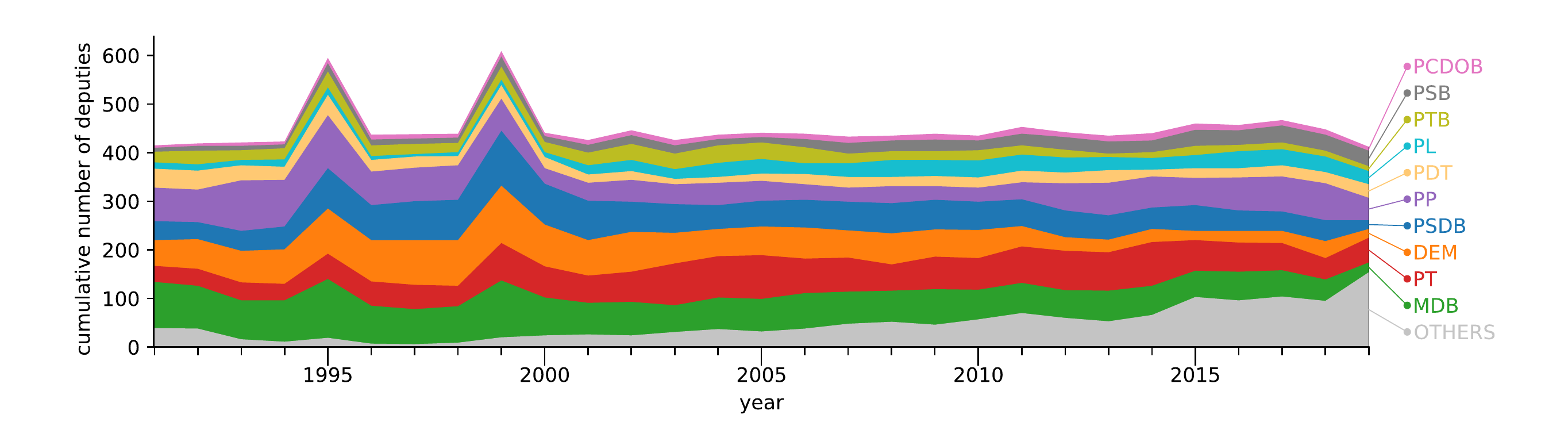}
    \caption{Number of seats of political parties as observed in the Brazilian Chamber of Deputies between 1991 and 2019. The number of deputies from nontraditional parties (``others'') clearly increasing along time. Only the deputies in the largest component of the network were considered.
    }
    \label{fig:deps}
\end{figure}

The total number of propositions voted in each year is shown in Figure \ref{fig:props}. In Figure \ref{fig:deps}, we show the distribution of parties (i.e. the total number of parliamentarians) along time. Interestingly, the representation of nontraditional parties have been increasing along time, while the size of traditional political parties (e.g. MDB and PSDB) is decreasing. A visualization of the obtained network over time is shown in Figure~\ref{fig:visualization}.

%


\section{Results and Discussion} \label{sec:results}


{Our main objective here is to leverage network science to better understand the dynamics of groups of deputies from a networked perspective. Using the proposed methodology, we intend to address the following general research questions, which are analyzed in the specific case of the Brazilian Chamber of Deputies.}
\begin{enumerate}

    \item Is the organization in parties consistent with clusters formed via voting behavior?

    \item Does the diversity of political parties correspond to the diversity of clusters?

    \item Is it possible to observe and/or quantify political concepts such as fragmentation and approximation of political parties along time?

    \item Is there any signal from network topology indicating major political transitions, such as in presidential impeachment?

\end{enumerate}
The response to the above questions is important to understand the dynamics of deputies relationships, both at the microscopic and mesoscopic level. Such an better understanding may point out to better ways to improve the democratic system of a country. In the same line, the extraction of information from unstructured date might assist the population to take decisions on who they choose as representatives based on the voting history of deputies and parties alike.

In order to address the above research questions, we divide our analysis in two parts. In Section \ref{sec:gl}, we analyze the global properties of the networks and compare the groups defined via topology and political parties. In Section \ref{eq:meso}, we analyze the network at the mesoscopic (party) level.

\subsection{Global analysis} \label{sec:gl}

\begin{figure}[hbtp!]
\includegraphics[width=0.95\linewidth]{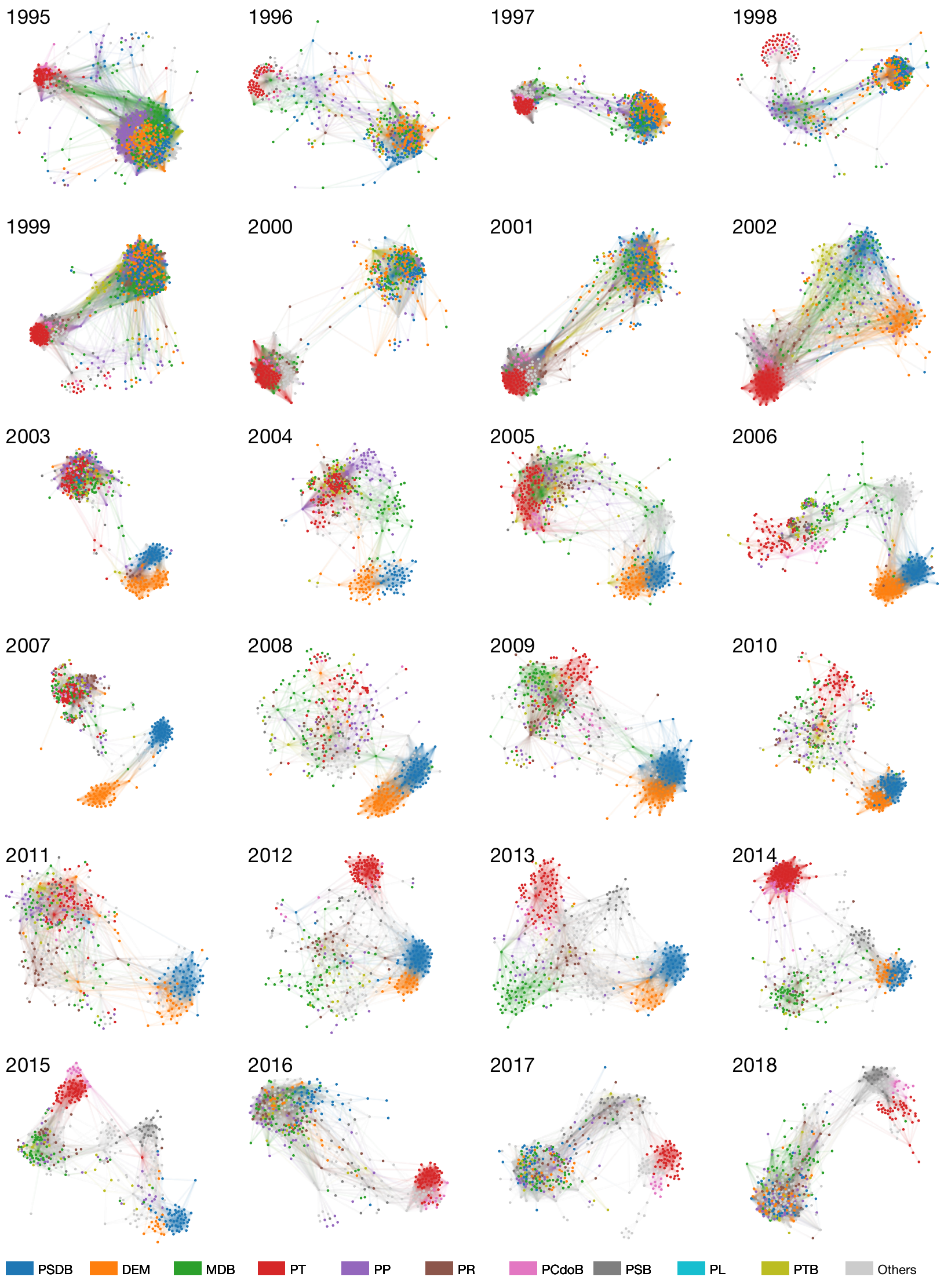}
    \caption{Visualization of the time evolving network obtained from voting data. Each yearly snapshot was obtained considering only the proposals voted in that year. Colors correspond to parties.}
    \label{fig:eeeevo}
\end{figure}

\begin{figure}[hbtp!]
    \centering
    \includegraphics[width=0.70\linewidth]{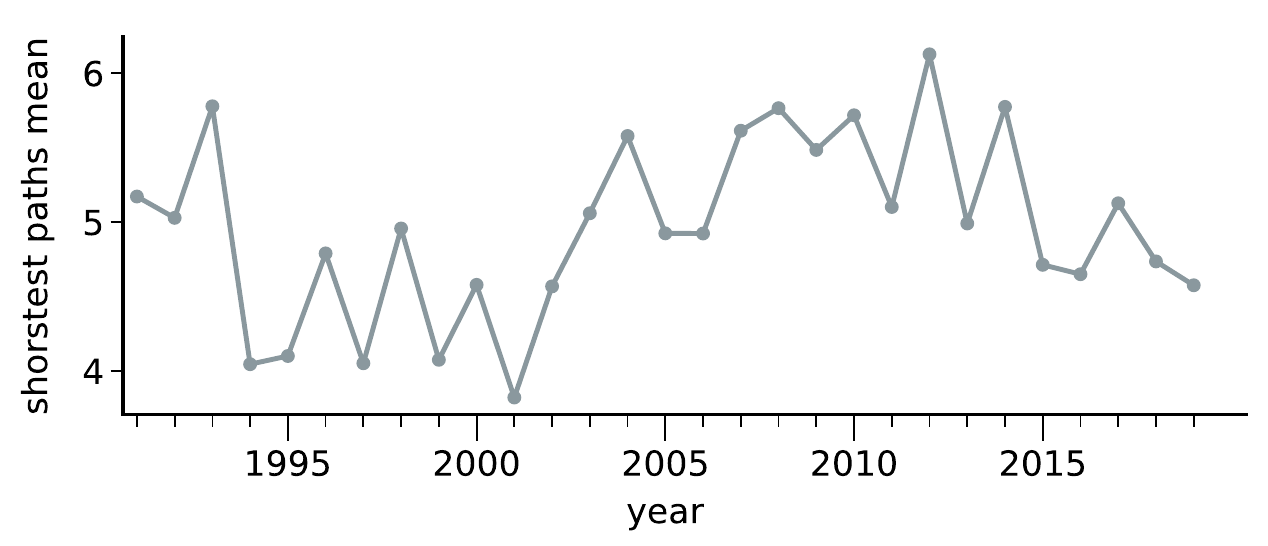}
    \caption{Evolution of the average shortest path length. While the average distance ranges in the interval  $4 \leq d_G \leq 6$, the networks obtained in the interval between 1994 and 2002 are typically more cohesive than the ones obtained between 2003 and 2012. 
    }
    \label{fig:short}
\end{figure}

The global evolution of the obtained political networks is depicted in Figure \ref{fig:eeeevo}, where colors represent political parties. An interactive visualization of these networks is available online~\footnote{\url{https://filipinascimento.github.io/politiciannetworks}}.
The first important property that can be recovered from the networks is the global \emph{cohesiveness}. This can be computed using a measurement that quantifies the distance between two nodes (deputies) in the network. If deputies vote in a similar fashion, one expects that the average distance between them is short. The obtained evolution of the global distances ($d_G$ in equation \ref{eq:distglob}) is shown in Figure \ref{fig:short}. All distances typically range in the interval $4 \leq d_G \leq 6$. Two patterns, however, can be identified. In the period between 1994 and 2002, distances are in the interval $4 \leq d_G \leq 5$. Conversely, in the period between 2003 and 2014, the average distances typically are typically higher, in the interval $5 \leq d_G \leq 6$. Interestingly these two intervals correspond to opposite parties occupying the presidency. This observation might indicate that different government ideologies may affect the way in which deputies vote.


\begin{figure}[hbtp!]
    \centering
    \includegraphics[width=0.80\linewidth]{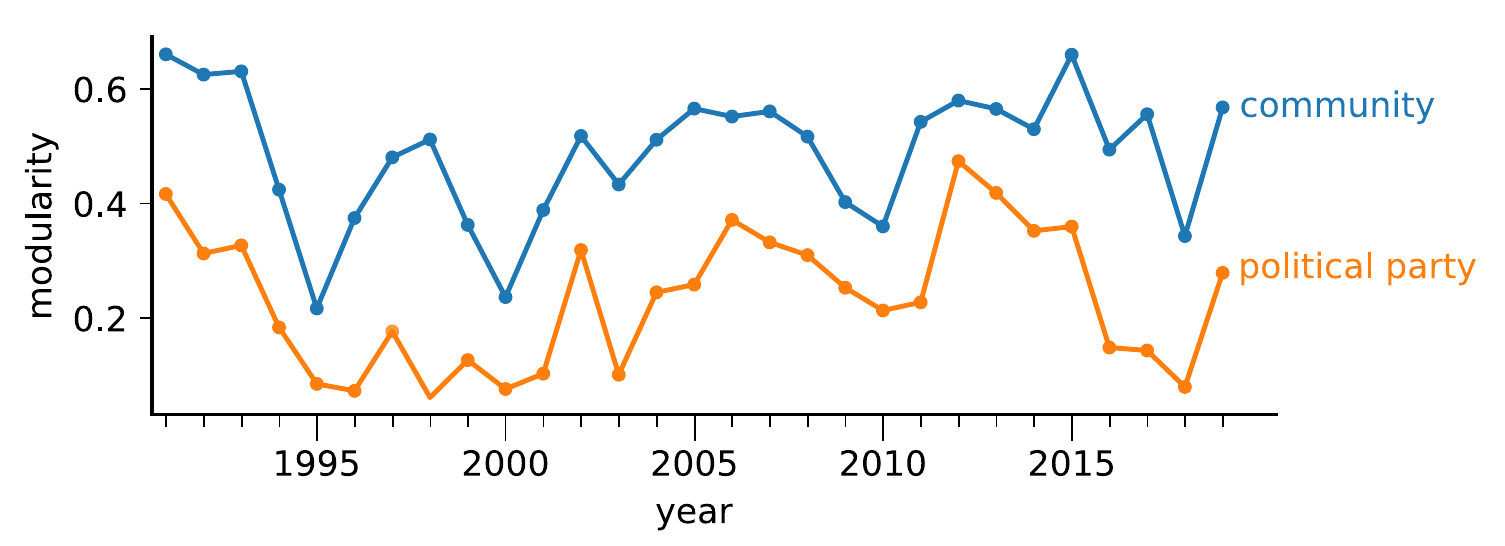}
    \caption{Evolution of the modularity measurement. To compute the modularity, we considered groups obtained from the community detection method (blue curve) and clusters matching deputies' political parties (orange curve). }
    \label{fig:modparty}
\end{figure}

The \emph{consistency} of political parties can be measured by probing whether  groups defined via political parties are consistent with topological groups, which are obtained using only the connectivity of the network. Topological groups are obtained via modularity optimization and, for this reason, this method tends to cluster nodes in the same group if they are strongly connected in the group, with a few external links. The evolution of the quality of the clustering obtained via topology and political parties are shown in Figure \ref{fig:modparty}. The modularity of groups obtained via topology outperforms the modularity of groups obtained via political parties. This means that there is a high tendency of deputies of different political parties to be connected. While the difference in modularity seems to be roughly constant along time, it is possible to observe higher deviations in particular periods. This is apparent e.g. in the years 2016 and 2017, which coincides with a period of post-impeachment political transition in Brazil. In both years, only PT was found to form an independent community, according to Figure \ref{fig:eeeevo}. An opposite effect occurred in 2015, when the difference in modularity reached its minimum value. This is evident in Figure \ref{fig:eeeevo}, because groups formed by parties are consistent with topological clusters.

\begin{figure}[hbtp!]
    \centering
    \includegraphics[width=0.75\linewidth]{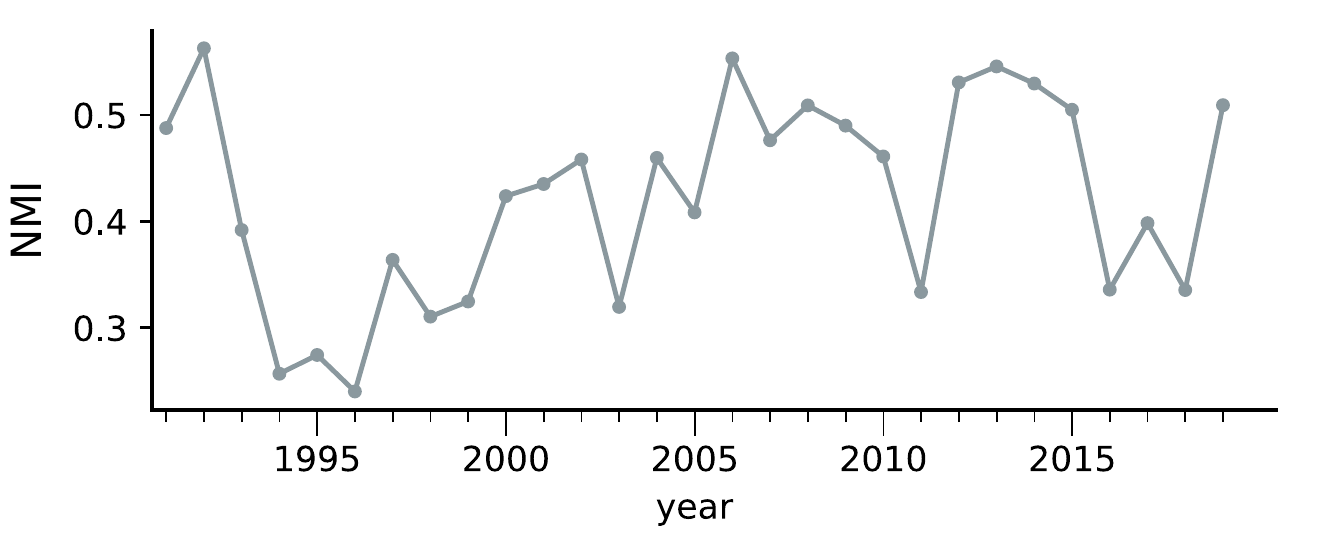}
    \caption{Normalized Mutual Information (NMI) between communities and political parties. A major transition affecting the NMI occurred just before the first PSDB presidency, from 1993 to 1994.}
    \label{fig:mutualinfo}
\end{figure}

The divergence between topological clusters and political parties can be further analyzed via Normalized Mutual Information (NMI)~\cite{zhang2015evaluating}, which measures the quality of the obtained partitions (topological communities) based on a set of nodes membership labels (political parties). The evolution of the NMI index is shown in Figure \ref{fig:mutualinfo}. Overall, the correspondence between political parties and topological partitions varies along time, but a major rupture seems to occurs just before the first PSDB term (1994) and just before impeachment (2016). Notably, the highest discrepancy (lowest value of NMI) occurred in 1996, which is in accordance with the visualization provided in Figure \ref{fig:eeeevo}.
This result confirms that, in the Brazilian political scenario, in most years the natural organization of deputies in communities has a low correspondence with party organization.

\begin{figure}[hbtp!]
    \centering
    \includegraphics[width=0.9\linewidth]{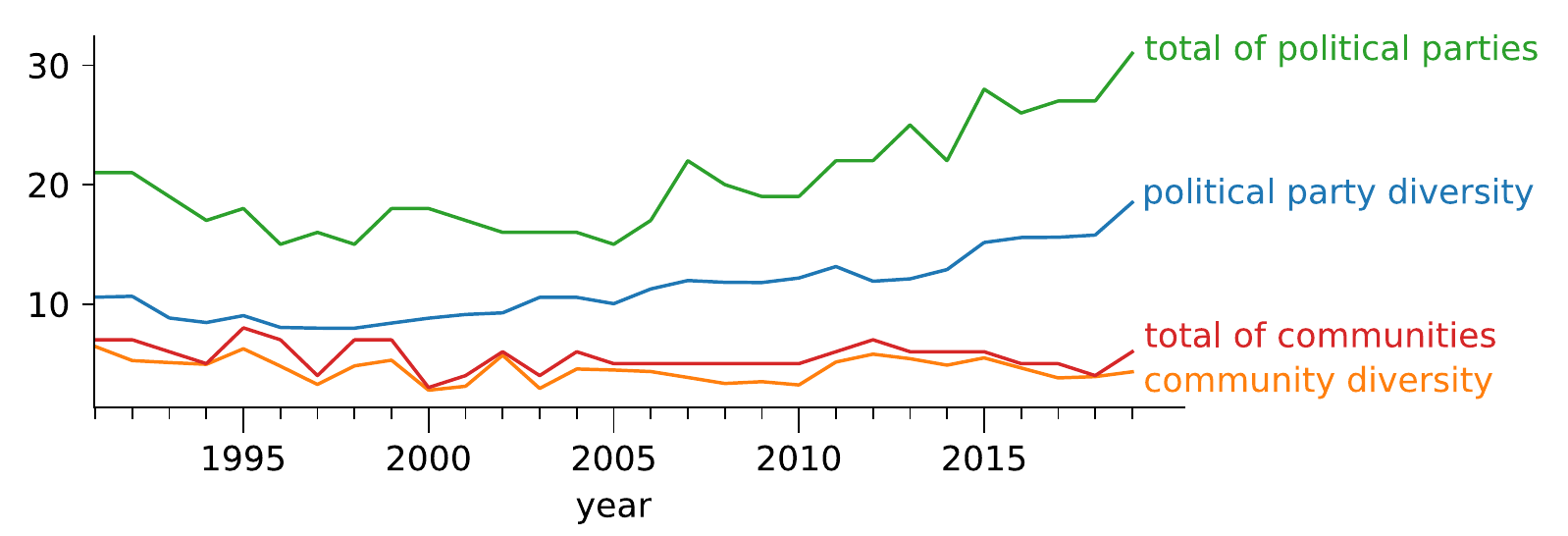}
    \caption{Evolution of clusters size and the diversity of clusters size. Even though the Brazilian Chamber of Deputies is represented by 29 political parties, the effective number of topological clusters (community diversity) in 2019 is only $3.37$.}
    \label{fig:divs}
\end{figure}

The discrepancy between political parties and topological clusters is also evident when one study the actual and effective quantity of groups.
The \emph{effective} number of parties and communities was computed using the true diversity of the distribution of clusters size, as described in Section \ref{submet3}. The results are depicted in Figure \ref{fig:divs}.
Since 2005, an increase in the total number of parties (green curve) is clear. In 2019, the Brazilian Chamber of Deputies is made up of roughly 29 political parties. When the effective number of parties is analyzed (blue curve), the number of parties decreases to 20 political parties. Surprisingly, the topology of the network revealed the existence of only about 5 clusters of deputies. In other words, even though 29 political parties exist, $D(\textrm{communities}) \simeq 3.37$ (see equation \ref{eq:dive}) in 2019. In other words, only 3.37 groups are effectively present in the Brazilian political scenario in 2019.


\subsection{Mesoscopic analysis} \label{eq:meso}

The obtained networks can also shed light into the understanding of the political dynamic at the mesoscopic level. An interesting analysis concerns the quantification of the pairwise distance between political parties along time.  This analysis is important, e.g. to quantify the level of \emph{coalision} (or \emph{opposition}) between parties and party \emph{isolation}. Such concepts are oftentimes used in politics because abrupt or recurrent changes in these quantities may precede a relevant change in the political scenario.

\begin{figure}[hbtp!]
    \centering
    \includegraphics[width=\linewidth]{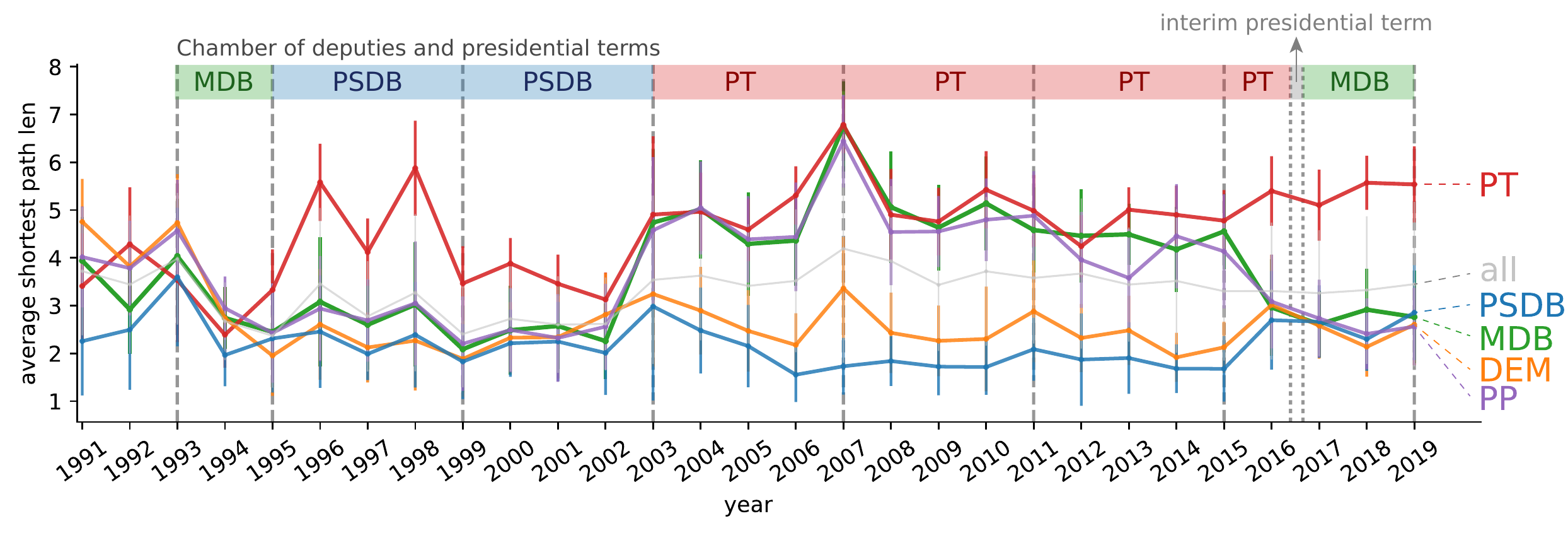}
    \caption{Average shortest path length between PSDB deputies and other parties. Distances are computed using equation \ref{eq:distab}.
    The PSDB fragmentation (i.e. $d(\textrm{PSDB},\textrm{PSDB})$) is computed using equation \ref{eq:frag}. Presidential terms are highlight in the period between 1993 and the end of 2018.}
    \label{fig:psdb}
\end{figure}

In Figure \ref{fig:psdb}, we show the temporal evolution of the distances between PSDB and other parties. The distances are computed using equation \ref{eq:distab}. In this figure, {we also show presidential terms in different colors, including the transition post-impeachment led by MDB presidency ($2016-2018$, green background)}. Interestingly, along the years of PSDB presidency, PT was found to be the most distant party from PSDB. This is consistent with the traditional sentiment of polarization between PT and PSDB in these years~\footnote{\url{https://en.wikipedia.org/wiki/Brazilian_Social_Democracy_Party}}~\footnote{\url{https://en.wikipedia.org/wiki/Workers\%27_Party_(Brazil)}}. During PT presidency, we note that PT and PSDB were still distant, however, two additional parties (MDB and PP) were found to increase (and keep) a larger distance from PSDB. This is consistent with the real political scenario, since MDB and PT shared a \emph{coalition} along the period characterized by PT presidency. These observations are consistent with the visualization provided in Figure \ref{fig:eeeevo}.

Another interesting pattern that can be extracted from Figure \ref{fig:psdb} concerns the evolution of PSDB fragmentation (i.e. $F(\textrm{PSDB}) = d(\textrm{PSDB},\textrm{PSDB})$), as defined in equation \ref{eq:frag}. The PSDB fragmentation corresponds to the blue curve. Note that the typical distance between PSDB deputies is compatible with the distance between PSDB and DEM virtually along all considered years and, markedly, before 2002. Such a proximity between these two parties is also evident from the visualization of the network structure. A strong concordance with MDB and PP is also apparent before 2002 and after 2016.

\begin{figure}[hbtp!]
    \centering
    \includegraphics[width=\linewidth]{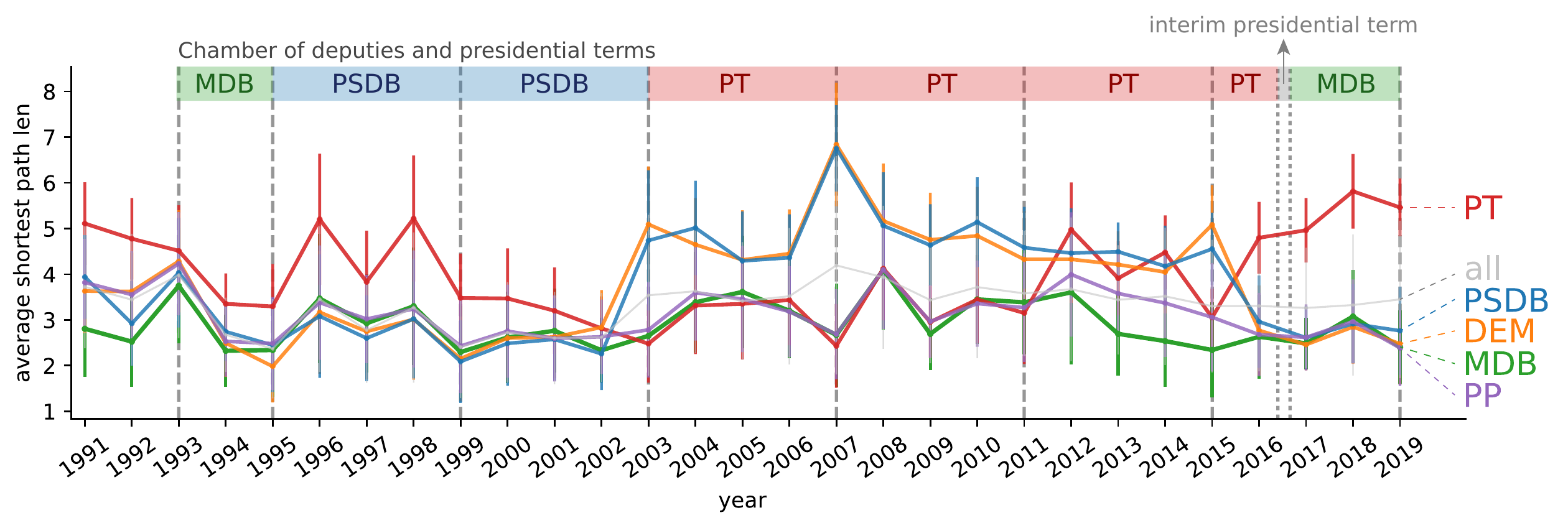}
    \caption{Average shortest path length between MDB deputies and other parties. Distances are computed using equation \ref{eq:distab}. MDB fragmentation (i.e. $d(\textrm{MDB},\textrm{MDB})$ is computed using equation \ref{eq:frag}. 
    }
    \label{fig:pmdb}
\end{figure}

\begin{figure}[hbtp!]
    \centering
    \includegraphics[width=\linewidth]{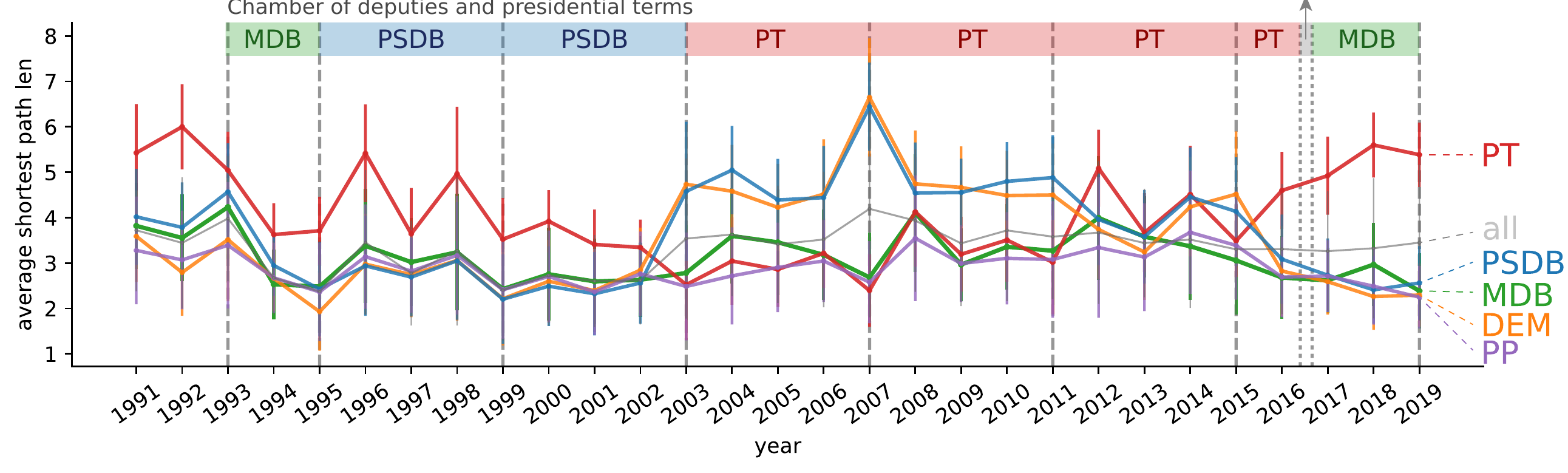}
    \caption{Average shortest path length between PP deputies and other parties. Distances are computed using equation \ref{eq:distab}. The PP fragmentation (i.e. $d(\textrm{PP},\textrm{PP})$ is computed using equation \ref{eq:frag}. Note that the distances behavior is similar with the ones found for MDB (see Figure \ref{fig:pmdb}).}
    \label{fig:pp}
\end{figure}

In Figure \ref{fig:pmdb}, we show the distances between MDB and other parties. In this case, the transition in government seems to play a prominent role in the behavior of this party. Between 1991 and 2001, the only party distant from MDB is PT. Conversely, between 2002 and 2011, PT and MDB became much closer. This is a consequence of the the political coalition shared by these parties along these years. Differently from the traditional alliance in previous years, MDB dissociated itself from PSDB and DEM between 2002 and 2016 (see also Figure \ref{fig:eeeevo}). The rapprochement between these parties took place only in 2016, in the post-impeachment transition.  The MDB fragmentation along time is roughly constant, however, apart from 2012-2015, the fragmentation is always similar to the typical distance between MDB and the closest parties. Finally, it is interesting to note that MDB and PP display a similar evolution of distances, according to Figures \ref{fig:pmdb} and \ref{fig:pp}, respectively.

\begin{figure}[hbtp!]
    \centering
    \includegraphics[width=\linewidth]{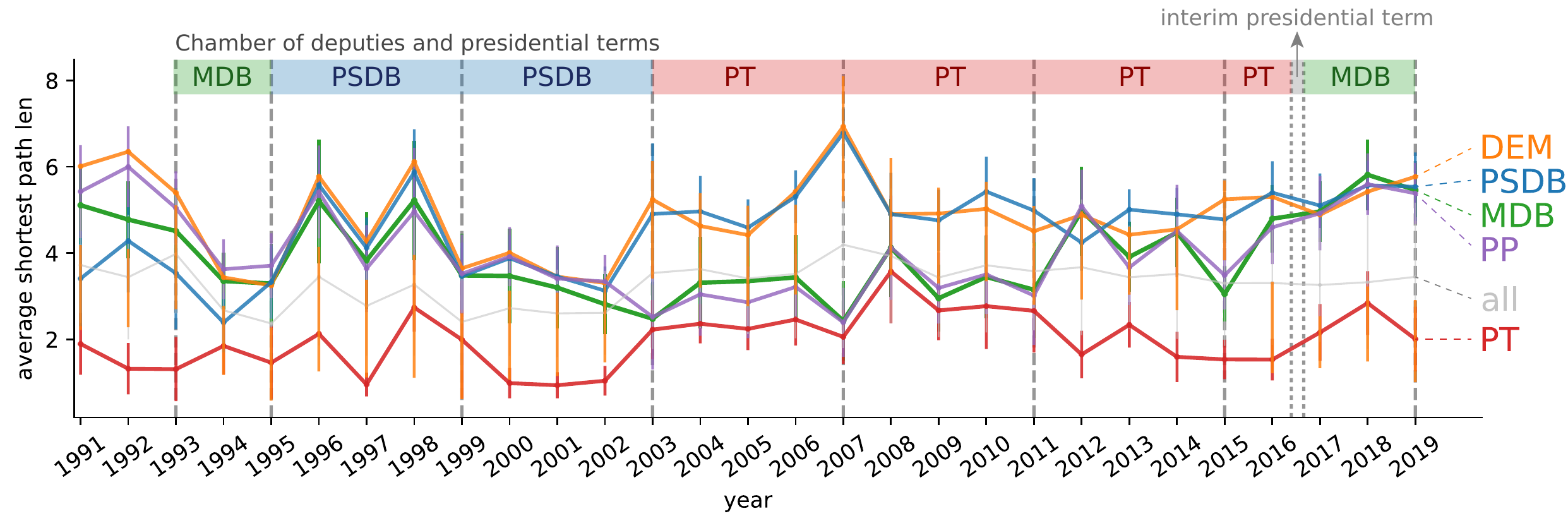}
    \caption{Average shortest path length between PT deputies and other parties. Distances are computed using equation \ref{eq:distab}. The PT fragmentation (i.e. $d(\textrm{PT},\textrm{PT})$) is computed using equation \ref{eq:frag}.}
    \label{fig:pt}
\end{figure}

The distance between PT and other political parties is depicted in Figure \ref{fig:pt}. It is clear from the data the PT and PSDB are continuously distant from each other since 1995. During PSDB presidency, the isolation of PT is clear: no other party is closer to PT than PT itself. In the first years of PT presidency, PT became closer to other parties, including MDB and PP. Some years before a new transition in governments, however, PT became distant from all others, specially after 2015. This isolation scenario became consistent even after the presidential impeachment (2016).
%

\begin{figure}[hbtp!]
    \centering
    \includegraphics[width=1\linewidth]{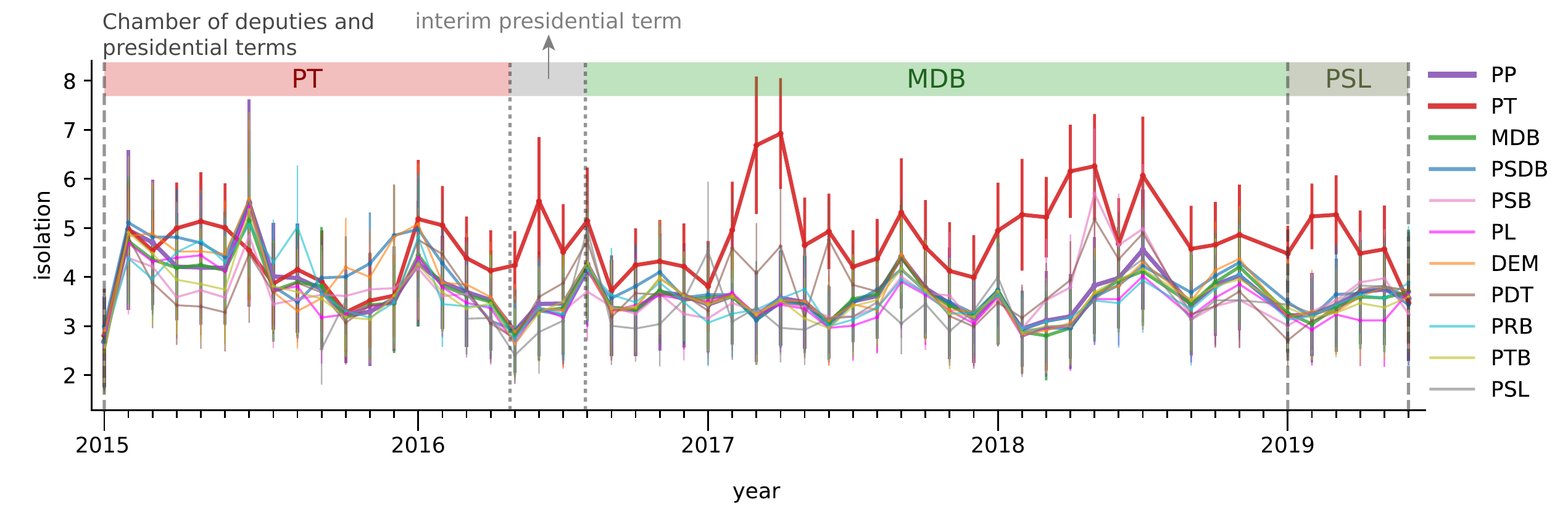}
    \caption{Isolation of political parties few months before and after impeachment. The isolation of parties is computed using equation \ref{eq:iso}.
    PT turned out to be the most isolated party a few months before impeachment. The relative isolation became stronger after 2017.}
    \label{fig:isolaimp}
\end{figure}

In Figure \ref{fig:isolaimp}, we analyze the \emph{isolation} of political parties few months before and after the impeachment of the Brazilian President, Dilma Rousseff (PT). PT and MDB terms are represented in red and green background colors, respectively. The transition between these two presidencies is shown in gray background color. Apart from PT, all parties have similar values and patterns of isolation along time.  While PT isolation was compatible with many the behavior displayed by many other parties, its isolation increased (relatively to other parties) months before impeachment. During the first months of the interim presidential term, PT isolation increased relatively to other parties. After 2017, PT turned out to be the most isolated political party. PT isolation is also clear in the visualization provided in Figure \ref{fig:eeeevo}, specially in 2017.
A sudden change in relative isolation months before impeachment could indicate the vulnerability of the party, and potentially a signal of a major change in the political scenario.

\section{Conclusion} \label{sec:conc}

In a democratic nation, the political organization can reflect its cultural context, economic and social interactions. The understanding of the role of deputies and parties alike is essential to improve any democratic system. Here we proposed a framework to assist the understanding of the dynamics of political parties. The relationship between deputies in the lower chamber was represented as a network, where two deputies are linked if they voted in a similar way in different propositions. Using concepts borrowed from network science, we defined politics concepts such as \emph{isolation}, \emph{fragmentation} and \emph{coalition}, which were measured in terms of topological distances.

Several interesting results were found when the proposed framework was applied to analyze the Brazilian Chamber of Deputies. During PSDB term, PT was found to be the most distant party from PSDB and vice-versa. The coalition shared by parties was easily identified from both distance metrics and the visualization of the obtained networks. The networks displayed a modular topology, however, we found  a low degree of consistency when comparing topological groups and political parties in most years. Surprisingly, we found that even though the Brazilian political scenario is highly fragmented, only a few topological groups emerge. While in 2019 the lower chamber comprised 29 distinct parties, we found that \emph{effectively} there are roughly only 3 clusters of deputies, according to the diversity measurement. Finally, a detailed analysis in the period between 2015 and 2019 revealed that PT isolation increased a few months before the presidential impeachment of Dilma Rousseff (PT). This could be an early sign before impeachment proceedings against former president Rousseff effectively started.

As future work, we intend to use additional information to analyze and possibly predict the results of votes based on the history of deputies, parties. This could be accomplished by using textual data obtained from the propositions and transcripts from deputies' discourses.
While the focus of this manuscript was in identifying and quantifying relevant political quantities via network science, we advocate that this framework could also be used in other scenarios. An analogous approach could be used to analyze other types of
social networks with similar characteristics, i.e. entities with pre-defined labels that interact
in a complex manner. This is the case of the \emph{Science of Science} field~\cite{fortunato2018science}, where authors interact via collaborations and, at the same time, they can be labeled according to diverse criteria: country, university, research field/group or others.



\section*{Acknowledgments}

\noindent
DRA acknowledges financial support from São Paulo Research Foundation (FAPESP Grant no. 16/19069-9) and CNPq-Brazil (Grant no. 304026/2018-2). ACMB acknowledges Capes-Brazil for sponsorship. 

\newpage

\bibliographystyle{ieeetr}

\end{document}